# Sequences, Bent Functions and Jacobsthal sums

Tor Helleseth and Alexander Kholosha

The Selmer Center
Department of Informatics, University of Bergen
P.O. Box 7800, N-5020 Bergen, Norway
{Tor.Helleseth,Alexander.Kholosha}@uib.no

**Abstract.** The $p$-ary function $f(x)$ mapping $\mathrm{GF}(p^{4k})$ to $\mathrm{GF}(p)$ and given by $f(x) = \mathrm{Tr}_{4k}\left(ax^d + bx^2\right)$ with $a, b \in \mathrm{GF}(p^{4k})$ and $d = p^{3k} + p^{2k} - p^k + 1$ is studied with the respect to its exponential sum. In the case when either $a^{p^k(p^k+1)} \neq b^{p^k+1}$ or $a^2 = b^d$ with $b \neq 0$, this sum is shown to be three-valued and the values are determined. For the remaining cases, the value of the exponential sum is expressed using Jacobsthal sums of order $p^k + 1$. Finding the values and the distribution of those sums is a long-lasting open problem.

**Keywords:** Cyclotomic number, Jacobsthal sum, $p$-ary bent function, polynomial over finite field, Walsh transform.

## 1 Introduction

Niho in [1, Theorem 3-7] and Helleseth in [2] studied the cross correlation between two binary $m$-sequences that differ by the decimation $2^{3k} - 2^{2k} + 2^k + 1$. They proved that the cross-correlation function is four-valued and found the distribution. In [3], Helleseth and Kholosha constructed a $p$-ary weakly regular binomial bent function that has an exponent of this type in its first term (the second term is a square). This gave the infinite class of nonquadratic generalized bent functions built over the fields of an arbitrary odd characteristic. In this paper, we take $n = 4k$, an odd prime $p$ and examine $p$-ary functions having the form $f(x) = \mathrm{Tr}_n\left(ax^d + bx^2\right)$ with $a, b, x \in \mathrm{GF}(p^n)$ and $d = p^{3k} + p^{2k} - p^k + 1$. Functions of this type with $a$ and $b$ being nonzero belong to the class of *binomials*. Note that $d$ is cyclotomic equivalent to the Niho exponent (with 2 changed to $p$) and that $\gcd(d, p^n - 1) = 2$ since $d = (p^{2k} - 1)(p^k + 1) + 2$.

Given a function $f(x)$ mapping $\mathrm{GF}(p^n)$ to $\mathrm{GF}(p)$, the direct and inverse *Walsh transform* operations on $f$ are defined at a point by the following respective identities:

$$S_f(y) = \sum_{x \in \mathrm{GF}(p^n)} \omega^{f(x) - \mathrm{Tr}_n(yx)} \quad \text{and} \quad \omega^{f(x)} = \frac{1}{p^n} \sum_{y \in \mathrm{GF}(p^n)} S_f(y) \omega^{\mathrm{Tr}_n(yx)}$$

* This work was supported by the Norwegian Research Council and partially by the grant NIL-I-004 from Iceland, Liechtenstein and Norway through the EEA and Norwegian Financial Mechanisms.

where $\mathrm{Tr}_n() : \mathrm{GF}(p^n) \to \mathrm{GF}(p)$ denotes the absolute trace function, $\omega = e^{\frac{2\pi i}{p}}$ is the complex primitive $p^{\mathrm{th}}$ root of unity and elements of $\mathrm{GF}(p)$ are considered as integers modulo $p$.

According to [4], $f(x)$ is called a *p-ary bent function* (or *generalized bent function*) if all its Walsh coefficients satisfy $|S_f(y)|^2 = p^n$. A bent function $f(x)$ is called *regular* (see [4, Definition 3] and [5, p. 576]) if for every $y \in \mathrm{GF}(p^n)$ the normalized Walsh coefficient $p^{-n/2}S_f(y)$ is equal to a complex $p^{\mathrm{th}}$ root of unity, i.e., $p^{-n/2}S_f(y) = \omega^{f^*(y)}$ for some function $f^*$ mapping $\mathrm{GF}(p^n)$ into $\mathrm{GF}(p)$. A bent function $f(x)$ is called *weakly regular* if there exists a complex $u$ having unit magnitude such that $up^{-n/2}S_f(y) = \omega^{f^*(y)}$ for all $y \in \mathrm{GF}(p^n)$. Recently, weakly regular bent functions were shown to be useful for constructing certain combinatorial objects such as partial difference sets, strongly regular graphs and association schemes (see [6, 7]). This justifies why the classes of (weakly) regular bent functions are of independent interest. For a comprehensive reference on monomial and quadratic $p$-ary bent functions we refer reader to [8].

It is known that taking $f(x)$ with $a = b = 1$, results in a weakly regular bent function and the exact value of its Walsh transform coefficients (and value distribution) can be found as follows.

**Theorem 1 ([3]).** *Let $n = 4k$. Then $p$-ary function $f(x)$ mapping $\mathrm{GF}(p^n)$ to $\mathrm{GF}(p)$ and given by*

$$f(x) = \mathrm{Tr}_n\left(x^{p^{3k}+p^{2k}-p^k+1} + x^2\right)$$

*is a weakly regular bent function. Moreover, for $y \in \mathrm{GF}(p^n)$ the corresponding Walsh transform coefficient of $f(x)$ is equal to*

$$S_f(y) = -p^{2k}\omega^{\mathrm{Tr}_k(x_0)/4} \ ,$$

*where $x_0$ is a unique root in $\mathrm{GF}(p^k)$ of the polynomial*

$$y^{p^{2k}+1} + (y^2+X)^{(p^{2k}+1)/2} + y^{p^k(p^{2k}+1)} + (y^2+X)^{p^k(p^{2k}+1)/2} \ .$$

*In particular, if $y^2 \in \mathrm{GF}(p^{2k})$ then $x_0 = -\mathrm{Tr}_k^{2k}(y^2)$. Also, every value $-p^{2k}\omega^i$ with $i = \{1, \ldots, p-1\}$ occurs $p^{2k-1}(p^{2k}+1)$ times in the Walsh spectrum of $f(x)$ and $-p^{2k}$ occurs $(p^{2k-1}-1)(p^{2k}+1)+1$ times.*

The general case when $a, b \in \mathrm{GF}(p^n)$ is much more complicated. It seems to be hard to find the Walsh transform coefficients of $f(x)$ at an arbitrary point, so here we calculate the exponential sum of $f(x)$, i.e., $S_f(0)$. This is equal to the cross-correlation function between two sequences of length $(p^n-1)/2$ obtained by the decimation of an $m$-sequence by $d$ and 2 or can be seen as a codeword weight in the corresponding $p$-ary linear code. We relate this value to the number of zeros, a particular polynomial has in a cyclic subgroup of order $p^{2k}+1$ of the multiplicative group of $\mathrm{GF}(p^n)$. Moreover, we show that if either $a^{p^k(p^k+1)} \neq b^{p^k+1}$ or $a^2 = b^d$ with $b \neq 0$ then the exponential sum of $f(x)$ is three-valued. Some steps towards finding the distribution of these values are made but the exact distribution still remains an open problem. For the remaining options for choosing $(a,b)$, we show that $S_f(0)$ can be expressed using the Jacobsthal sums of order $p^k+1$ and the number of possible values grows with $k$. In Section 2, we compute the cyclotomic numbers of order $p^k+1$ in $\mathrm{GF}(p^{2k})^*$ (this was already done in [9] but our proof is direct). These are used to prove few important properties that should facilitate finding the value distribution in general.

## 2 Cyclotomic Numbers of Order $p^k+1$

Let $\nu$ be a primitive element of $\mathrm{GF}(p^{2k})$ and let $C_t$ $(t=0,\dots,p^k)$ denote the *cyclotomic classes* of order $p^k+1$ in the multiplicative group of $\mathrm{GF}(p^{2k})$, i.e., $C_t=\{\nu^{(p^k+1)i+t} \mid i=0,\dots,p^k-2\}$. The number of elements $x\in C_i$ such that $x+1\in C_j$ is called the *cyclotomic number* and denoted $(i,j)$. Since $-1\in C_0$ in our case, we can also take $x-1$ in the definition of the cyclotomic numbers. Note that since the cyclotomic numbers of order $p^k+1$ are *uniform* (see [9]), their values can easily be determined. Nevertheless, in the following lemma, we give a direct and easy proof using the technique suggested for the binary case in [10, Sec. 5].

**Lemma 1.** *For any $i,j=0,\dots,p^k$, the cyclotomic numbers of order $p^k+1$ in $\mathrm{GF}(p^{2k})$ are*

$$(i,j)=\begin{cases}1, & \textit{if} \quad i\neq j \textit{ and } ij\neq 0\\ p^k-2, & \textit{if} \quad i=j=0\\ 0, & \textit{otherwise}\end{cases} .$$

*Proof.* Note that $\mathrm{GF}(p^{2k})^*=\bigcup_{t=0}^{p^k} C_t$ and $-1=\nu^{(p^{2k}-1)/2}\in C_0$.

$$\begin{aligned}p^{2k}(i,j) &= \sum_{z\in \mathrm{GF}(p^{2k})}\sum_{x\in C_i}\sum_{y\in C_j}\omega^{\mathrm{Tr}_{2k}(z(x-y-1))}\\ &=(p^k-1)^2+\sum_{t=0}^{p^k}\sum_{z\in C_t}\omega^{-\mathrm{Tr}_{2k}(z)}\sum_{x\in C_i}\omega^{\mathrm{Tr}_{2k}(zx)}\sum_{y\in C_j}\omega^{-\mathrm{Tr}_{2k}(zy)}\\ &=(p^k-1)^2+\sum_{t=0}^{p^k}P_tP_{t+i}P_{t+j}\ ,\end{aligned}$$

where indices of $P_t$ are calculated modulo $p^k+1$ and

$$\begin{aligned}P_t &= \sum_{x\in C_t}\omega^{\mathrm{Tr}_{2k}(x)}=\frac{1}{p^k+1}\sum_{z\in \mathrm{GF}(p^{2k})^*}\omega^{\mathrm{Tr}_{2k}\left(z^{p^k+1}\nu^t\right)}\\ &\overset{(*)}{=}\begin{cases}p^k-1, \text{ if} & t=(p^k+1)/2\\ -1, & \text{otherwise}\end{cases}\end{aligned}$$

for $t=0,\dots,p^k$ and $(*)$ follows from [8, Lemma 2 (iii)]. Therefore, if $i\neq j$ and $ij\neq 0$ then

$$(i,j)=p^{-2k}\left((p^k-1)^2+3(p^k-1)-(p^k-2)\right)=1\ .$$

Similarly, it is easy to see that $(0,0) = p^k - 2$ and in the rest of the cases, $(i,j) = 0$. $\square$

## 3 Estimate of the Jacobsthal Sums of Order $p^k + 1$

Following [11, Definition 5.49], for any $a \in \mathrm{GF}(p^{2k})^*$, define a *Jacobsthal sum* of order $n$ as

$$H_n(a) = \sum_{x \in \mathrm{GF}(p^{2k})} \eta(x^{n+1} + ax) \ ,$$

where $\eta(\cdot)$ is the quadratic character of $\mathrm{GF}(p^{2k})$ extended by setting $\eta(0) = 0$. Define also a companion sum

$$I_n(a) = \sum_{x \in \mathrm{GF}(p^{2k})^*} \eta(x^n + a) \ .$$

It is well known (see, e.g., [11, Theorem 5.50]) that $I_{2n}(a) = I_n(a) + H_n(a)$.

Take $n = p^k + 1$, any $a \in \mathrm{GF}(p^{2k}) \setminus \mathrm{GF}(p^k)$ and assume $a^{-1} \in C_i$. Then $i \neq 0$ since $C_0 = \mathrm{GF}(p^k)^*$, and we can compute

$$\begin{aligned}
I_{p^k+1}(a) &= \eta(a) \sum_{x \in \mathrm{GF}(p^{2k})^*} \eta\big(x^{p^k+1}/a + 1\big) \\
&= (-1)^i (p^k+1) \left( \sum_{j=0}^{(p^k-1)/2} (i, 2j) - \sum_{j=0}^{(p^k-1)/2} (i, 2j+1) \right) \\
&\overset{(*)}{=} (p^k+1) \begin{cases} \frac{p^k+1}{2} - 2 - \frac{p^k+1}{2} = -2, & \text{if } i \text{ is even} \\ -\frac{p^k+1}{2} + 1 + \frac{p^k+1}{2} - 1 = 0, & \text{if } i \text{ is odd} \end{cases} \\
&= -(p^k+1)(\eta(a) + 1) \ , \qquad (1)
\end{aligned}$$

where $(*)$ follows from Lemma 1. Note that $\eta(a) = (-1)^{p^k+1-i} = (-1)^i$ since $a \in C_{p^k+1-i}$. Calculating $H_{p^k+1}(a)$ (that is equivalent to calculating $I_{2(p^k+1)}(a)$) is not that easy. In the following theorem, we provide an estimate for these values. Note that this estimate is much better than the one in [11, p. 233] which becomes trivial if $n = p^k + 1$. Computations show that the bound found in Theorem 2 is achievable.

**Theorem 2.** *For any* $a \in \mathrm{GF}(p^{2k}) \setminus \mathrm{GF}(p^k)$,

$$|H_{p^k+1}(a)| \leq 2p^{k/2}(p^k + 1) \ .$$

*Proof.* Since $I_{2(p^k+1)}(a) = I_{p^k+1}(a) + H_{p^k+1}(a)$ and the exact value of $I_{p^k+1}(a)$ was found in (1), we need to estimate $I_{2(p^k+1)}(a)$. Raising elements of $\mathrm{GF}(p^{2k})^*$ to the power of $p^k + 1$ defines a $(p^k + 1)$-to-1 mapping onto $\mathrm{GF}(p^k)^*$. Thus, denoting $y = x^{p^k+1}$, we obtain from the definition

$$\frac{I_{2(p^k+1)}(a)}{p^k+1} + \eta(a) = \sum_{y \in \mathrm{GF}(p^k)} \eta(y^2 + a) = N(a) - p^k \ ,$$

where $N(a)$ is the number of pairs $(y,t) \in \mathrm{GF}(p^k) \times \mathrm{GF}(p^{2k})^*$ that satisfy $y^2 + a = t^2$.

If $\mu$ is a primitive element of $\mathrm{GF}(p^k)$ then $\mu^{1/2} \in \mathrm{GF}(p^{2k}) \setminus \mathrm{GF}(p^k)$ and any element $x \in \mathrm{GF}(p^{2k})$ has a unique representation as $x = x_0 + 2\mu^{1/2}x_1$ with $x_0, x_1 \in \mathrm{GF}(p^k)$. This way, assume $a = a_0 + 2\mu^{1/2}a_1$ and $t = t_0 + 2\mu^{1/2}t_1$. Thus, $y^2 + a = t^2$ is equivalent to $y^2 + a_0 = t_0^2 + 4\mu t_1^2$ with $a_1 = 2t_0t_1$. Note that $t_1 \neq 0$ since in the opposite case, $t \in \mathrm{GF}(p^k)$ that leads to $a \in \mathrm{GF}(p^k)$. Combining the latter equations we obtain $y^2t_0^2 + a_0t_0^2 = t_0^4 + \mu a_1^2$. Therefore, $N(a)$ is equal to the number of pairs $(y,z) \in \mathrm{GF}(p^k) \times \mathrm{GF}(p^k)^*$ that satisfy

$$y^2z^2 + Az^2 = z^4 + C \ ,$$

where $C = \mu a_1^2 \neq 0$ and $A = a_0$, both in $\mathrm{GF}(p^k)$.

Now we can calculate

$$\begin{aligned}
&2p^k\left(N(a) - p^k + 1\right) = 2 \sum_{y,z,l \in \mathrm{GF}(p^k);\, zl \neq 0} \omega^{\mathrm{Tr}_k\left(l(z^4 - Az^2 + C) - ly^2z^2\right)} \\
&= \sum_{zl \neq 0} \omega^{\mathrm{Tr}_k\left(l^2(z^4 - Az^2 + C)\right)} \sum_y \omega^{-\mathrm{Tr}_k\left(l^2z^2y^2\right)} \\
&\quad + \sum_{zl \neq 0} \omega^{\mathrm{Tr}_k\left(\mu l^2(z^4 - Az^2 + C)\right)} \sum_y \omega^{-\mathrm{Tr}_k\left(\mu l^2z^2y^2\right)} \\
&= \sum_y \omega^{-\mathrm{Tr}_k\left(y^2\right)} \left( \sum_{z \neq 0,\, l} \omega^{\mathrm{Tr}_k\left(l^2(z^4 - Az^2 + C)\right)} - \sum_{z \neq 0,\, l} \omega^{\mathrm{Tr}_k\left(\mu l^2(z^4 - Az^2 + C)\right)} \right) \\
&= 2 \sum_y \omega^{-\mathrm{Tr}_k\left(y^2\right)} \sum_{z^5 - Az^3 + Cz \neq 0,\, l} \omega^{\mathrm{Tr}_k\left(l^2(z^4 - Az^2 + C)\right)} \\
&= 2p^k s\zeta(-1) \sum_{z \neq 0} \zeta(z^4 - Az^2 + C) \\
&= 2p^k \sum_{z \neq 0} (1 + \zeta(z))\zeta(z^2 - Az + C) \\
&= 2p^k \sum_z \zeta(z^2 - Az + C) - \zeta(C) + \sum_z \zeta(z^3 - Az^2 + Cz) \\
&\overset{(*)}{=} 2p^k \sum_z \zeta(z^3 - Az^2 + Cz) \ ,
\end{aligned}$$

where $\zeta(\cdot)$ is the quadratic character of $\mathrm{GF}(p^k)$ extended by setting $\zeta(0) = 0$; $s = (-1)^k$ if $p \equiv 3 \pmod 4$ and $s = 1$ otherwise; and $(*)$ follows from [11, Theorem 5.48] since $z^2 - Az + C$ can not have both roots in $\mathrm{GF}(p^k)$ equal ($C$ is a nonsquare in $\mathrm{GF}(p^k)$). Also note that $s\zeta(-1) \equiv 1$. Thus,

$$\frac{I_{2(p^k+1)}(a)}{p^k + 1} = \sum_z \zeta(z^3 - a_0z^2 + \mu a_1^2 z) - \eta(a) - 1 = N - p^k - \eta(a) - 1 \quad \text{and}$$

$$\frac{H_{p^k+1}(a)}{p^k + 1} = N - p^k \ ,$$

where $N$ denotes the number of points on the elliptic curve $f^2 = z^3 - Az^2 + Cz$ over $\mathrm{GF}(p^k)$ *excluding* the point at infinity. It remains to use Hasse theorem [12, p. 138] giving $|N - p^k| \leq 2p^{k/2}$ to obtain the claimed result (also, [11, Theorem 5.41] can be used). $\square$

## 4 Calculating the Exponential Sum of $f(x)$

In this section, we consider the function $f(x)$ with arbitrary coefficients $a, b \in \mathrm{GF}(p^n)$. If $n$ is even, let $U$ denote a cyclic subgroup of order $p^{n/2}+1$ of the multiplicative group of $\mathrm{GF}(p^n)$ (generated by $\xi^{p^{n/2}-1}$, where $\xi$ is a primitive element of $\mathrm{GF}(p^n)$).

**Theorem 3.** *Let $n = 4k$. For any $a, b \in \mathrm{GF}(p^n)$, define the following $p$-ary function mapping $\mathrm{GF}(p^n)$ to $\mathrm{GF}(p)$*

$$f(x) = \mathrm{Tr}_n\left(ax^{p^{3k}+p^{2k}-p^k+1} + bx^2\right) \ .$$

*Then the Walsh transform coefficient of $f(x)$ evaluated at point zero is equal to*

$$S_f(0) = p^{2k}(2N(a,b) - 1) \ ,$$

*where $2N(a,b)$ is the number of zeros in $U$ of the polynomial*

$$L(X) = b^{p^{2k}}X + aX^{p^k} + bX^{p^{2k}} + a^{p^{2k}}X^{p^{3k}} \ . \tag{2}$$

*Proof.* Let $\xi$ be a primitive element of $\mathrm{GF}(p^n)$ and also denote $d = p^{3k} + p^{2k} - p^k + 1$. If we let $x = \xi^j y^{p^{2k}+1}$ for $j = 0, \dots, p^{2k}$ and $y$ running through $\mathrm{GF}(p^n)^*$ then $x$ will run through $\mathrm{GF}(p^n)^*$ in total $p^{2k}+1$ times. Also note that $d - 2 = (p^{2k}-1)(p^k+1)$ and thus, $d(p^{2k}+1) \equiv 2(p^{2k}+1) \pmod{p^n - 1}$. Therefore, the Walsh transform coefficient of $f(x)$ evaluated at point zero is equal to

$$\begin{aligned}
S_f(0) - 1 &= \sum_{x \in \mathrm{GF}(p^n)^*} \omega^{\mathrm{Tr}_n\left(ax^{p^{3k}+p^{2k}-p^k+1}+bx^2\right)} \\
&= \frac{1}{p^{2k}+1}\sum_{j=0}^{p^{2k}} \sum_{y \in \mathrm{GF}(p^n)^*} \omega^{\mathrm{Tr}_n\left(a\xi^{dj}y^{2(p^{2k}+1)}+b\xi^{2j}y^{2(p^{2k}+1)}\right)} \\
&= \sum_{j=0}^{p^{2k}} \sum_{z \in \mathrm{GF}(p^{2k})^*} \omega^{\mathrm{Tr}_n\left(\left(a\xi^{dj}+b\xi^{2j}\right)z^2\right)} \\
&= \sum_{j=0}^{p^{2k}} \sum_{z \in \mathrm{GF}(p^{2k})^*} \omega^{\mathrm{Tr}_{2k}\left(\xi^{(p^{2k}+1)j}L(\xi^{(p^{2k}-1)j})z^2\right)} \\
&\overset{(*)}{=} \sum_{j=0}^{p^{2k}} I\big(L(\xi^{(p^{2k}-1)j}) \neq 0\big)\left(-sp^k\eta\left(\xi^{(p^{2k}+1)j}L\big(\xi^{(p^{2k}-1)j}\big)\right) - 1\right) \\
&\qquad + 2N(a,b)(p^{2k}-1) \ ,
\end{aligned}$$

where $z = y^{p^{2k}+1} \in \mathrm{GF}(p^{2k})^*$ is a $(p^{2k}+1)$-to-1 mapping of $\mathrm{GF}(p^n)^*$, $(*)$ is obtained by [8, Corollary 3], $s = (-1)^k$ if $p \equiv 3 \pmod 4$ and $s = 1$ otherwise, $I(\cdot)$ is the indicator function, $\eta(\cdot)$ is the quadratic character of $\mathrm{GF}(p^{2k})$ and since

$$\begin{aligned}
&\mathrm{Tr}_{2k}^{n}\left(a\xi^{dj} + b\xi^{2j}\right) = a\xi^{dj} + b\xi^{2j} + a^{p^{2k}}\xi^{djp^{2k}} + b^{p^{2k}}\xi^{2jp^{2k}} \\
&= \xi^{(p^{2k}+1)j}\left(a\xi^{p^k(p^{2k}-1)j} + b\xi^{-(p^{2k}-1)j} + a^{p^{2k}}\xi^{-p^k(p^{2k}-1)j} + b^{p^{2k}}\xi^{(p^{2k}-1)j}\right) \\
&= \xi^{(p^{2k}+1)j}L\big(\xi^{(p^{2k}-1)j}\big).
\end{aligned}$$

also noting that $\xi^{-(p^{2k}-1)j} = \xi^{p^{2k}(p^{2k}-1)j}$.

Further, note that for any $j = 0, \ldots, \frac{p^{2k}-1}{2}$ with $L\big(\xi^{(p^{2k}-1)j}\big) \neq 0$ we have

$$\eta\left(\xi^{(p^{2k}+1)(j+(p^{2k}+1)/2)}L\big(\xi^{(p^{2k}-1)(j+(p^{2k}+1)/2)}\big)\right) = -\eta\left(\xi^{(p^{2k}+1)j}L\big(\xi^{(p^{2k}-1)j}\big)\right)$$

since $L(-x) = -L(x)$ for any $x \in \mathrm{GF}(p^n)$ and $\eta(-1) = \eta\Big(\big(\xi^{p^{2k}+1}\big)^{(p^{2k}-1)/2}\Big) = 1$. Therefore,

$$S_f(0) = -(p^{2k} + 1 - 2N(a,b)) + 2N(a,b)(p^{2k} - 1) + 1 = p^{2k}(2N(a,b) - 1) \ .$$

Obviously, the number of zeros in $U$ of $L(X)$ is even since $-U = U$ and $L(-x) = -L(x)$ for any $x \in \mathrm{GF}(p^n)$. □

In the following corollary, we prove that it is sufficient to consider just two inequivalent cases, when $b$ is a square and nonsquare in $\mathrm{GF}(p^n)^*$, for instance, taking $b = 1$ and $b = \xi$, where $\xi$ is a primitive element of $\mathrm{GF}(p^n)$.

**Corollary 1.** *Under the conditions and using the notations of Theorem 1, for any $h \in \mathrm{GF}(p^n)^*$,*

$$N(a,b) = N(ah^d, bh^2) \ .$$

*Proof.* Recalling definition (2), $2N(ah^d, bh^2)$ is equal to the number of zeros in $U$ of the polynomial

$$\begin{aligned}
(bh^2)^{p^{2k}}X + ah^dX^{p^k} &+ bh^2X^{p^{2k}} + (ah^d)^{p^{2k}}X^{p^{3k}} \\
&= h^{p^{2k}+1}\Big(b^{p^{2k}}h^{p^{2k}-1}X + ah^{p^k(p^{2k}-1)}X^{p^k} \\
&\quad + bh^{-(p^{2k}-1)}X^{p^{2k}} + a^{p^{2k}}h^{-p^k(p^{2k}-1)}X^{p^{3k}}\Big) \\
&= h^{p^{2k}+1}\left(b^{p^{2k}}Y + aY^{p^k} + bY^{p^{2k}} + a^{p^{2k}}Y^{p^{3k}}\right) \ ,
\end{aligned}$$

where $Y = h^{p^{2k}-1}X$ and since $h^{p^{2k}-1} \in U$. By definition, the latter polynomial has $2N(a,b)$ zeros in $U$. □

In what follows, we consider separately the cases when either $a^{p^k(p^k+1)} \neq b^{p^k+1}$ or $a^2 = b^d$ with $b \neq 0$; and when $a^{p^k(p^k+1)} = b^{p^k+1}$ with $a^2 \neq b^d$, where $d = p^{3k} + p^{2k} - p^k + 1$. This covers all the value space for the pairs $(a,b) \neq (0,0)$.

### 4.1 Case $a^{p^k(p^k+1)} \neq b^{p^k+1}$

In this subsection, we show that the exponential sum of $f(x)$ takes on just three values $-p^{2k}$, $p^{2k}$ and $3p^{2k}$ when either $a^{p^k(p^k+1)} \neq b^{p^k+1}$ or $a^2 = b^d$ with $b \neq 0$.

**Proposition 1.** *Let $n = 4k$ and take any $a, b \in \mathrm{GF}(p^n)$ such that either $a^2 = b^d$ with $b \neq 0$ or $a^{p^k(p^k+1)} \neq b^{p^k+1}$. Then polynomial $L(X)$ defined in (2) has none, two or four zeros in $U$, i.e., $N(a, b) \in \{0, 1, 2\}$. Moreover, if $a^{p^k(p^k+1)} \neq b^{p^k+1}$ then zeros of $L(X)$ in $\mathrm{GF}(p^n)$ are the same as of*

$$F(X) = \left(a^{p^k(p^k+1)} - b^{p^k+1}\right)X^{p^{2k}} + \left(a^{p^{2k}} b^{p^{3k}} - ab^{p^k}\right)X^{p^k} + \left(a^{p^k(p^k+1)} - b^{p^k+1}\right)^{p^k} X \, .$$

*Proof.* First, assume $a^2 = b^d \neq 0$ with $a^{p^k(p^k+1)} = b^{p^k+1}$. Then

$$a^{p^k(p^k+1)} = b^{dp^k(p^k+1)/2} = b^{p^k\left(p^n-1+2(p^{3k}+1)\right)/2} = b^{(p^n-1)/2} b^{p^k+1} = b^{p^k+1} \quad (3)$$

if and only if $b$ is a square in $\mathrm{GF}(p^n)^*$. By Corollary 1, taking $h = b^{-1/2}$ we obtain that $N(a, b) = N(ab^{-d/2}, 1) = N(\pm 1, 1)$. By definition, $2N(\pm 1, 1)$ is equal to the number of zeros in $U$ of $x \pm x^{p^k} + x^{-1} \pm x^{-p^k}$. For any $v \in U$ we obtain

$$v \pm v^{p^k} + v^{-1} \pm v^{-p^k} = v^{-(p^k+1)}\left(v^{p^k+1} \pm 1\right)\left(v \pm v^{p^k}\right) = 0$$

only if $v^{2(p^k+1)} = 1$ or $v^{2(p^k-1)} = 1$ which leads to $v^2 = 1$ since $\gcd(2(p^k + 1), p^{2k}+1) = \gcd(2(p^k-1), p^{2k}+1) = 2$. Thus, $v = \pm 1$ that gives no zeros when $a = b^{d/2}$ and two when $a = -b^{d/2}$.

From now on assume $a^{p^k(p^k+1)} \neq b^{p^k+1}$. Note that zeros of

$$a^{p^{2k}} L(X)^{p^k} - b^{p^k} L(X) = F(X)$$

are exactly the union of solution sets for $L(X) = 0$ and $a^{p^{2k}} L(X)^{p^k-1} = b^{p^k}$. Since $L(x) \in \mathrm{GF}(p^{2k})$ for any $x \in \mathrm{GF}(p^n)$ and assuming $L(x) \neq 0$, the latter equation can have solution only if $a^{p^{2k}(p^k+1)} = b^{p^k(p^k+1)}$ that is equivalent to $a^{p^k(p^k+1)} = b^{p^k+1}$. Thus, $L(X)$ and $F(X)$ have the same zeros. Also note that $F(x)$ degenerates if and only if $a^{p^k(p^k+1)} = b^{p^k+1}$ since in this case,

$$\begin{aligned} a^{p^{2k}} b^{p^{3k}} - ab^{p^k} &= a^{-p^k}\left(a^{p^k(p^k+1)} b^{p^{3k}} - (a^{p^k(p^k+1)})^{p^{3k}} b^{p^k}\right) \\ &= a^{-p^k}\left(b^{p^{3k}+p^k+1} - b^{p^{3k}+p^k+1}\right) = 0 \ , \end{aligned} \quad (4)$$

i.e., $ab^{p^k} \in \mathrm{GF}(p^{2k})$.

Raising the elements of $U$ to the power of $p^k - 1$ defines a 2-to-1 mapping onto $U_+$ the set of squares of $U$ since $\gcd(p^k - 1, p^{2k} + 1) = 2$. Thus, making a substitution $Y = X^{p^k-1}$ and denoting $A = a^{p^k(p^k+1)} - b^{p^k+1}$ we obtain the polynomial

$$P(Y) = AY^{p^k+1} + \left(a^{p^{2k}} b^{p^{3k}} - ab^{p^k}\right)Y + A^{p^k}$$

that has $N(a,b)$ zeros in $U_+$. Further, assuming $Y^{p^{2k}} = Y^{-1}$, we obtain

$$
\begin{aligned}
&AY^2P(Y)^{p^k} - \left(a^{p^{3k}}b - a^{p^k}b^{p^{2k}}\right)YP(Y) - A^{p^k}P(Y) \\
&= A^{p^k}\left(A^{p^{3k}}Y^2 - \left(a^{p^{2k}}b^{p^{3k}} - ab^{p^k} + a^{p^{3k}}b - a^{p^k}b^{p^{2k}}\right)Y - A^{p^k}\right) \ .
\end{aligned}
$$

Since $A \neq 0$, the latter polynomial is non-degenerate and has at most two zeros in $\mathrm{GF}(p^n)$ which also means that $N(a,b) \leq 2$. □

### 4.2 Case $a^{p^k(p^k+1)} = b^{p^k+1}$ and Jacobsthal Sums

In this subsection, we consider the case when $a^{p^k(p^k+1)} = b^{p^k+1}$ with $a^2 \neq b^d$ and express the exponential sum of $f(x)$ using Jacobsthal sums of order $p^k+1$.

**Proposition 2.** *Let $n = 4k$ and take any $a, b \in \mathrm{GF}(p^n)$ such that $a^{p^k(p^k+1)} = b^{p^k+1}$ and $a^2 \neq b^d$. If $2N(a,b)$ is the number of zeros in $U$ of the polynomial $L(x)$ defined in (2) then*

$$
N(a,b) = \#\left\{c \in \mathrm{GF}(p^k) \mid (cg)^2 - b^{p^{2k}+1} \text{ is a nonsquare in } \mathrm{GF}(p^{2k})\right\} \ , \quad (5)
$$

*where $g$ is any element in $\mathrm{GF}(p^{2k})^*$ with $g^{p^k-1} = -b^{p^{3k}}/a$.*

*Proof.* Note that in our case, $a, b \neq 0$ and $g \in \mathrm{GF}(p^{2k})^*$ since $\left(b^{p^{3k}}/a\right)^{p^k+1} = 1$. Take any $u \in U$ with $L(u) = 0$. Multiplying both sides of $L(u) = 0$ by $b^{p^{3k}}$ and using (4), we obtain

$$
a\left(b^{p^{2k}}u + bu^{-1}\right)^{p^k} + b^{p^{3k}}\left(b^{p^{2k}}u + bu^{-1}\right) = 0 \ . \quad (6)
$$

Denote $b^{p^{2k}}u + bu^{-1} = g \in \mathrm{GF}(p^{2k})$. Find solutions in $U$ of the quadratic equation $b^{p^{2k}}x + bx^{-1} = g$ which discriminant is equal to $D = g^2 - 4b^{p^{2k}+1} \in \mathrm{GF}(p^{2k})$.

First, assume $D$ is a square in $\mathrm{GF}(p^{2k})$. Then $u = (g \pm \sqrt{D})/2b^{p^{2k}}$ and $b^{p^{2k}}u \in \mathrm{GF}(p^{2k})^*$ resulting in $g = 2b^{p^{2k}}u \neq 0$ and $D = 0$. In this case, (6) is reduced to $au^{p^k-1} = -b^{p^{2k}}$. We also obtain that

$$
\left(b^{p^{2k}}u\right)^{p^{2k}+1} = b^{p^{2k}+1} = \left(b^{p^{2k}}u\right)^2
$$

that is equivalent to $b = u^2b^{p^{2k}}$ Then $u = \pm b^{-(p^{2k}-1)/2}$ and

$$
au^{p^k-1}b^{-p^{2k}} = ab^{-d/2} = -1
$$

that leads to $a = -b^{d/2}$. Thus, no solutions in $U$ exist if $a^2 \neq b^d$.

If $D$ is a nonsquare in $\mathrm{GF}(p^{2k})$ then there exists some $d \in \mathrm{GF}(p^n) \setminus \mathrm{GF}(p^{2k})$ such that $g^2 - 4b^{p^{2k}+1} = d^2$. Raising both sides of the latter identity to the power of $p^{2k}$, we obtain $g^2 - 4b^{p^{2k}+1} = d^{2p^{2k}} = d^2$ that leads to $d^{p^{2k}} = -d$ since $d \notin \mathrm{GF}(p^{2k})$. Solutions of $b^{p^{2k}}x + bx^{-1} = g$ are $x_{1,2} = (g \pm d)/2b^{p^{2k}}$ and

$$x_{1,2}^{p^{2k}+1} = \frac{g^{p^{2k}+1} \pm g^{p^{2k}} d \pm g d^{p^{2k}} + d^{p^{2k}+1}}{4b^{p^{2k}+1}} = \frac{g^2 \pm gd \mp gd - d^2}{4b^{p^{2k}+1}} = 1 \ .$$

Thus, $x_{1,2} \in U$.

Summarizing the arguments presented above, we conclude that if $a^{p^k(p^k+1)} = b^{p^k+1}$ and $a^2 \neq b^d$ then for any $g \in \mathrm{GF}(p^{2k})$, the equation $b^{p^{2k}}x + bx^{-1} = g$ has no solutions in $U$ if $g^2 - 4b^{p^{2k}+1}$ is a square in $\mathrm{GF}(p^{2k})$ and has two solutions in $U$ otherwise.

If $b^{p^{2k}}u + bu^{-1} \neq 0$ then (6) can be written as

$$\left(b^{p^{2k}}u + bu^{-1}\right)^{p^k-1} = -\frac{b^{p^{3k}}}{a} \ .$$

Raising elements of $\mathrm{GF}(p^{2k})^*$ to the power of $p^k - 1$ defines a $(p^k - 1)$-to-1 mapping onto the cyclic subgroup of $\mathrm{GF}(p^{2k})^*$ of order $p^k + 1$ and all elements in the set $\{2cg \mid c \in \mathrm{GF}(p^k)^*\}$ with $g^{p^k-1} = -b^{p^{3k}}/a$ map to the same element $-b^{p^{3k}}/a$. Include $c = 0$ to take care of the case when $b^{p^{2k}}u + bu^{-1} = 0$. The discriminant of the quadratic equation $b^{p^{2k}}x + bx^{-1} = 2cg$, equal to $(2cg)^2 - 4b^{p^{2k}+1}$, is a square if and only if $D = (cg)^2 - b^{p^{2k}+1}$ is a square. Only those $c \in \mathrm{GF}(p^k)$ with $D$ being a nonsquare contribute two solutions to $2N(a,b)$. $\square$

Like in Proposition 2, assume $a, b \in \mathrm{GF}(p^n)^*$ with $a^2 \neq b^d$ and $g \in \mathrm{GF}(p^{2k})^*$ with $g^{p^k-1} = -b^{p^{3k}}/a$. In this case, $b^{p^{2k}+1}/g^2 \notin \mathrm{GF}(p^k)$ since

$$\left(\frac{b^{p^{2k}+1}}{g^2}\right)^{p^k-1} = \frac{a^2 b^{(p^{2k}+1)(p^k-1)}}{b^{2p^{3k}}} = \frac{a^2}{b^d} \neq 1 \tag{7}$$

which also means that $(cg)^2 - b^{p^{2k}+1} \neq 0$ for any $c \in \mathrm{GF}(p^k)$. Therefore, by Proposition 2,

$$\begin{aligned} p^k - 2N(a,b) &= \sum_{c \in \mathrm{GF}(p^k)} \eta\big((cg)^2 - b^{p^{2k}+1}\big) \\ &= \eta\big(-b^{p^{2k}+1}\big) + \frac{1}{p^k+1} \sum_{x \in \mathrm{GF}(p^{2k})^*} \eta\big(x^{2(p^k+1)} - b^{p^{2k}+1}/g^2\big) \\ &= \eta\big(-b^{p^{2k}+1}/g^2\big) + \frac{I_{2(p^k+1)}\big(-b^{p^{2k}+1}/g^2\big)}{p^k+1} \\ &\overset{(1)}{=} \frac{H_{p^k+1}\big(-b^{p^{2k}+1}/g^2\big)}{p^k+1} - 1 \ . \end{aligned}$$

We conclude that

$$2N(a,b) = p^k - \frac{H_{p^k+1}\big(-b^{p^{2k}+1}/g^2\big)}{p^k+1} + 1 \tag{8}$$

and, by Theorem 3,

$$S_f(0) = p^{2k}\left(p^k - \frac{H_{p^k+1}\big(-b^{p^{2k}+1}/g^2\big)}{p^k+1}\right) \ .$$

Thus, finding the value distribution of $S_f(0)$ when $a^{p^k(p^k+1)} = b^{p^k+1}$ with $a^2 \neq b^d$, is related to finding the values of the Jacobsthal sum of order $p^k+1$. In the following corollary, we list some basic properties of $N(a,b)$.

**Corollary 2.** *Under the conditions of Proposition 2,*

*(i)* $N(a,b) = \begin{cases} N(a^{-1},b^{-1}), & \text{if } b^{(p^n-1)/2} = 1 \\ p^k+1-N(a^{-1},b^{-1}), & \text{otherwise} \end{cases}$ ;

*(ii)* $N(a,b) + N(-a,b) = N(a,b) + N(a,-b) = p^k+1$*;*

*(iii)* $N(-a,b) = \begin{cases} p^k+1-N(a^{-1},b^{-1}), & \text{if } b^{(p^n-1)/2} = 1 \\ N(a^{-1},b^{-1}), & \text{otherwise} \end{cases}$ ;

*(iv) if* $b^{(p^n-1)/2} = 1$ *(resp.* $b^{(p^n-1)/2} = -1$*) then* $N(a,b)$ *is an even (resp. odd) number;*

*(v)* $\left|N(a,b) - \frac{p^k+1}{2}\right| \leq p^{k/2}$*, in particular,* $N(a,b)$ *is positive and, if* $k > 2$ *then* $N(a,b) > 8$*;*

*(vi) if* $p \equiv -1 \pmod 4$*,* $k$ *is odd and* $b^{(p^n-1)/2} = 1$ *then* $N(a,b) = N(-a,b) = (p^k+1)/2$ *for* $a = \nu^{(p^{2k}-1)/4}b^{d/2}$*, where* $\nu$ *is a primitive element of* $\mathrm{GF}(p^{2k})$*;*

*(vii) for any* $b \in \mathrm{GF}(p^n)^*$*,*

$$\sum_{a\in \mathrm{GF}(p^n):\ a^{p^k(p^k+1)}=b^{p^k+1},\ a^2\neq b^d} N(a,b) = (p^k+1)\big(p^k - b^{(p^n-1)/2}\big)/2 \ . \quad (9)$$

*Proof.* First, note that $b^{p^{2k}+1}$ is a square in $\mathrm{GF}(p^{2k})$ if and only if $b^{(p^n-1)/2} = 1$, i.e., $b$ is a square in $\mathrm{GF}(p^n)$. Assume $c \neq 0$ in (5). Then if $b$ is a square (resp. nonsquare) in $\mathrm{GF}(p^n)$ then $(cg)^2 - b^{p^{2k}+1}$ is a nonsquare in $\mathrm{GF}(p^{2k})$ if and only if $(cg)^{-2} - b^{-(p^{2k}+1)}$ is a nonsquare (resp. square), since $-1 = \big(\nu^{(p^{2k}-1)/4}\big)^2$. If $b$ is a square in $\mathrm{GF}(p^n)$ then, by (5),

$$\begin{aligned} N(a,b) &= \#\left\{c \in \mathrm{GF}(p^k)^* \mid (cg)^2 - b^{p^{2k}+1} \text{ is a nonsquare in } \mathrm{GF}(p^{2k})\right\} \\ &= \#\left\{c \in \mathrm{GF}(p^k)^* \mid (cg^{-1})^2 - b^{-(p^{2k}+1)} \text{ is a nonsquare in } \mathrm{GF}(p^{2k})\right\} \\ &= N(a^{-1},b^{-1}) \end{aligned}$$

since $g^{-(p^k-1)} = -a/b^{p^{3k}}$ if $g^{p^k-1} = -b^{p^{3k}}/a$. Similarly, If $b$ is a nonsquare in $\mathrm{GF}(p^n)$ then

$$\begin{aligned} N(a,b) &= 1 + \#\left\{c \in \mathrm{GF}(p^k)^* \mid (cg)^2 - b^{p^{2k}+1} \text{ is a nonsquare in } \mathrm{GF}(p^{2k})\right\} \\ &= 1 + \#\left\{c \in \mathrm{GF}(p^k)^* \mid (cg^{-1})^2 - b^{-(p^{2k}+1)} \text{ is a square in } \mathrm{GF}(p^{2k})\right\} \\ &= 1 + p^k - 1 - \big(N(a^{-1},b^{-1}) - 1\big) \ . \end{aligned}$$

This proves (i).

For a pair $(a, b)$, the corresponding $g \in \mathrm{GF}(p^{2k})^*$ satisfies $g^{p^k-1} = -b^{p^{3k}}/a$. Then $\left(\nu^{(p^k+1)/2}g\right)^{p^k-1} = b^{p^{3k}}/a$ which means that $\nu^{(p^k+1)/2}g$ corresponds both to $(-a, b)$ and $(a, -b)$. Also, $\left(c\nu^{(p^k+1)/2}g\right)^2 = \nu^{p^k+1}c^2g^2$ and $\nu^{p^k+1}$ is a generator of $\mathrm{GF}(p^k)^*$. If $b^{p^{2k}+1}$ is a square in $\mathrm{GF}(p^{2k})$ then for any $c \in \mathrm{GF}(p^k)^*$, we have $cg^2/b^{p^{2k}+1} \in C_{2i}$ with $i \neq 0$ that follows from (7). In this case, by (5),

$$\begin{aligned} N(a,b) + N(-a,b) &= N(a,b) + N(a,-b) \\ &= 2\#\left\{c \in \mathrm{GF}(p^k)^* \mid cg^2 - b^{p^{2k}+1} \text{ is a nonsq. in } \mathrm{GF}(p^{2k})\right\} \\ &= 2\#\{x \in C_{2i} \mid x - 1 \text{ is a nonsquare in } \mathrm{GF}(p^{2k})\} \\ &= 2\sum_{j=0}^{(p^k-1)/2} (2i, 2j+1) \overset{(*)}{=} p^k + 1 \end{aligned}$$

since the set of nonsquares in $\mathrm{GF}(p^{2k})$ is equal to $\bigcup_{j=0}^{(p^k-1)/2} C_{2j+1}$ and where $(*)$ is obtained using Lemma 1. Similarly, if $b^{p^{2k}+1}$ is a nonsquare in $\mathrm{GF}(p^{2k})$ then for any $c \in \mathrm{GF}(p^k)^*$, we have $cg^2/b^{p^{2k}+1} \in C_{2i+1}$ and

$$\begin{aligned} N(a,b) + N(\pm a, \mp b) &= 2 + 2\#\{x \in C_{2i+1} \mid x - 1 \text{ is a square in } \mathrm{GF}(p^{2k})\} \\ &= 2 + 2\sum_{j=0}^{(p^k-1)/2} (2i+1, 2j) = p^k + 1 \end{aligned}$$

since the set of squares in $\mathrm{GF}(p^{2k})$ is equal to $\bigcup_{j=0}^{(p^k-1)/2} C_{2j}$. The additive term 2 comes from $c = 0$. This proves (ii), and (iii) follows directly by combining (i) and (ii).

If $b^{p^{2k}+1}$ is a square in $\mathrm{GF}(p^{2k})$ then $c \neq 0$ and $N(a, b)$ is even since $c^2$ is 2-to-1 on $\mathrm{GF}(p^k)^*$. If $b^{p^{2k}+1}$ is a nonsquare then $c = 0$ contributes 1 to $N(a, b)$ and makes it odd.

Combining (8) with Theorem 2 we immediately obtain the estimate in (v). Also note that $(p^k + 1)/2 - p^{k/2}$ grows both with $p$ and $k$. The lowest value is achieved when $p = 3$ and $k = 1$ giving $2 - 3^{1/2} > 0$ (thus, $N(a, b) > 0$) and if $k > 2$ then $N(a, b) \geq 14 - 27^{1/2} > 8$.

If $b$ is square in $\mathrm{GF}(p^n)^*$ then, by Corollary 1,

$$N(\pm\nu^{(p^{2k}-1)/4}b^{d/2}, b) = N(\pm\nu^{(p^{2k}-1)/4}, 1) \ .$$

Then (vi) follows from (ii) and (iii) since $a^{-1} = -a$ if and only if $a^2 = -1 = \nu^{(p^{2k}-1)/2}$ (we also have to remember the requirement $a^{p^k(p^k+1)} = b^{p^k+1}$ and $a^2 \neq b^d$ that in our case becomes $a^{p^k+1} = 1$ and $a \neq \pm 1$).

Take any $b \in \mathrm{GF}(p^n)^*$ and fix (conditions of Proposition 2 provide that $b \neq 0$). Note that $x^{p^k(p^k+1)} = b^{p^k+1}$ has $p^k + 1$ solutions in $\mathrm{GF}(p^n)$. If $b$ is a square in $\mathrm{GF}(p^n)$ then both $a = \pm b^{d/2}$ satisfy $a^{p^k(p^k+1)} = b^{p^k+1}$ (see (3)).

Thus, summation conditions in (9) are satisfied by $p^k - 1$ values of $a \in \mathrm{GF}(p^n)$. On the other hand, if $b$ is a nonsquare in $\mathrm{GF}(p^n)$ then $a^2 \neq b^{d/2}$ whenever $a^{p^k(p^k+1)} = b^{p^k+1}$. Therefore, (9) immediately follows from (ii). $\square$

Take any $b \in \mathrm{GF}(p^n)^*$ and fix. Having in mind Theorem 3 and Proposition 1, suppose $S_f(0)$ takes on the values $-p^{2k}$, $p^{2k}$ and $3p^{2k}$ respectively $r$, $s$ and $t$ times when $a \in \mathrm{GF}(p^n)$ and either $a^{p^k(p^k+1)} \neq b^{p^k+1}$ or $a^2 = b^d$ (actually, by Corollary 2 (iii), $S_f(0) \neq -p^{2k}$ in all the remaining cases).

First, assume $b$ is a square in $\mathrm{GF}(p^n)^*$. Then $r + s + t = p^n - p^k + 1$ (see the proof of Corollary 2 (vii)). Further, by Theorem 3,

$$\begin{aligned}\sum_{a \in \mathrm{GF}(p^n)} S_f(0) &= -rp^{2k} + sp^{2k} + 3tp^{2k} + p^{2k} \sum_{a^{p^k(p^k+1)} = b^{p^k+1},\ a^2 \neq b^d} (2N(a,b) - 1) \\ &\overset{(9)}{=} p^{2k}(-r + s + 3t) + p^{2k}(p^{2k} - 1 - p^k + 1) \\ &= p^{2k}(-r + s + 3t - p^k) + p^n \ .\end{aligned}$$

Similarly, if $b$ is a nonsquare in $\mathrm{GF}(p^n)^*$ then $r + s + t = p^n - p^k - 1$ and

$$\begin{aligned}\sum_{a \in \mathrm{GF}(p^n)} S_f(0) &\overset{(9)}{=} p^{2k}(-r + s + 3t) + p^{2k}((p^k + 1)^2 - p^k - 1) \\ &= p^{2k}(-r + s + 3t + p^k) + p^n \ .\end{aligned}$$

On the other hand, for any $b \in \mathrm{GF}(p^n)$,

$$\begin{aligned}\sum_{a \in \mathrm{GF}(p^n)} S_f(0) &= \sum_{a \in \mathrm{GF}(p^n)} \sum_{x \in \mathrm{GF}(p^n)} \omega^{\mathrm{Tr}_n\left(ax^{p^{3k}+p^{2k}-p^k+1} + bx^2\right)} \\ &= \sum_{x \in \mathrm{GF}(p^n)} \omega^{\mathrm{Tr}_n\left(bx^2\right)} \sum_{a \in \mathrm{GF}(p^n)} \omega^{\mathrm{Tr}_n\left(ax^{p^{3k}+p^{2k}-p^k+1}\right)} = p^n \ .\end{aligned}$$

Thus, if $b \neq 0$ then $r + s + t = p^n - p^k + b^{(p^n-1)/2}$ and $-r + s + 3t = b^{(p^n-1)/2}p^k$.

Note that finding the sum of squares of $S_f(0)$ is easy in our case. Therefore, knowing the values and the distribution of the Jacobsthal sum of order $p^k + 1$ would give us the third equation allowing to find $r$, $s$ and $t$. However, in this way we are facing some long-lasting open problems. On the other hand, it may be possible to extract some extra relations for the unknowns, thus, bypassing the problem of finding the value distribution of Jacobsthal sums. This is the first direct connection between sequences and Jacobsthal sums we are aware of. We find it interesting and believe that this gives an important link between sequences/codes and classical character sums.